\documentclass[10pt,a4paper]{article}

\usepackage{jucs2e}
\usepackage{url}
\usepackage{epsfig}
\usepackage{amssymb}
\usepackage{amsmath}
\usepackage{amsfonts}

\usepackage{xargs}                      
\usepackage{graphicx}
\usepackage{epstopdf}
\usepackage[colorinlistoftodos,prependcaption]{todonotes}

\usepackage[square]{natbib}
\usepackage{prettyref}
\newrefformat{sec}{[Section~\ref{#1}]}
\newrefformat{fig}{[Fig.~\ref{#1}]}
\newrefformat{tab}{[Tab.~\ref{#1}]}

\begin{document}

\title{Creativity in Mind: Evaluating and Maintaining Advances in Network Steganographic Research\footnote{This is a pre-print version. The final version will be available from \url{http://www.jucs.org/jucs_21} as an article of the Journal of Universal Computer Science (J.UCS), Vol.\ 21 (2015).}}

\author{{\bfseries Steffen Wendzel}\\
   (Department of Cyber Security, Fraunhofer FKIE, Bonn, Germany\\
   steffen.wendzel@fkie.fraunhofer.de)
   \and
   {\bfseries Carolin Palmer}\\
   (Justus-Liebig-Universit{\"a}t, Giessen, Germany \\
   carolin.palmer@psychol.uni-giessen.de)\\
}
\maketitle

\begin{abstract}
The research discipline of network steganography deals with the hiding of information within network transmissions, e.g. to transfer illicit information in networks with Internet censorship.
The last decades of research on network steganography led to more than hundred techniques for hiding data in network transmissions. However, previous research has shown that most of these hiding techniques are either based on the same idea or introduce limited novelty, enabling the application of existing countermeasures.
In this paper, we provide a link between the field of creativity and network steganographic research. We propose a framework and a metric to help evaluating the creativity bound to a given hiding technique. This way, we support two sides of the scientific peer review process as both authors and reviewers can use our framework to analyze the novelty and applicability of hiding techniques. At the same time, we contribute to a uniform terminology in network steganography.
\end{abstract}


\begin{category}
D.2.11, 
D.4.6,  
K.6.5,  
K.7.m   
\end{category}

\section{Introduction}

\emph{Steganography} is the research discipline that aims on hiding information within a medium; it dates back to ancient Greece, \citep[see][]{petitcolas1999information,Katzenbeisser:2000:IHT:555654}. For instance, information can be hidden in music scores, in paintings, on letters written using invisible ink, or on human skin, just to mention a few options \citep{petitcolas1999information}. Digital media steganography is a newer area of the research discipline and focuses on hiding information in digital content, such as audio files or digital images. The newest sub-discipline of steganography is \emph{network steganography}, which hides information within data sent over a computer network, such as Internet chats, network protocol headers, or within the content of \emph{Voice over IP} (VoIP) transmissions \citep{NetStegBook}. In comparison to cryptography, which aims on concealing the \emph{content} of a secret message, network steganography hides the \emph{existence} of a secret transmission. For this reason, network steganography is a dual-use good and can, for instance, be used by journalists to transfer illicit information when facing Internet censorship but it can also be applied by botnets to hide malicious data transfers.

In order to hide information within network transmissions, research led to more than hundred so-called \emph{hiding techniques}. These hiding techniques are functions that accept an input in form of data to be hidden. The hiding technique embeds the input into a transmission such that the input remains concealed within the transfer until a receiver applies another function to extract the embedded hidden information from the transmission.

So-called `patterns' are used in various sciences, from architecture to software engineering. Patterns are abstract descriptions of recurring designs, which provide a solution for a problem (in a given context). In our work, a pattern describes the design of a technique (solution), which hides data (problem) in network transmissions (context). In previous work \citep{Wendzel:CSUR}, it was shown that 109 network information hiding techniques developed between 1987 and 2013 can be reduced to only 11 different `hiding patterns', i.e. abstract descriptions of these hiding techniques.
Moreover, it was shown that the majority (approx.\ 70\%) of these 109 techniques is represented by only 4 hiding patterns. Although it must be noted that many of these techniques differ in slight detail, their novelty is limited. 
However, the amount of published techniques reflects a great demand of steganographic techniques and stresses the need for continuously improved solutions.

A major drawback of having a high number of similar hiding techniques is that countermeasures can be adjusted easily to slightly new hiding techniques 
while it would be harder to create completely new countermeasures which have to deal with entirely novel hiding techniques. For this reason, the research community should foster such fully novel methods.

Moreover, while the number of publications presenting hiding techniques is steadily increasing, the chances for redundant techniques with similar basis increase as well --- a problem which is also known by the patterns research community \citep{Henninger:PatternPracticesTR}. Another problem is that such redundancies lead to terminological inconsistencies as different terms can be used to describe similar (or equal) hiding techniques.
A means to handle redundancies and terminological inconsistencies is to compile surveys %
on a regular basis.

Another problem of published hiding techniques are the huge differences in the explanation of novelty and usefulness. To provide an example, some researchers motivate the quality of their hiding techniques on the basis of the channel capacity while others highlight the fact that their technique is the first to hide data inside a new network protocol header. The divergence of such provided arguments allows no comparison and hinders experimental verification by other peers.

In this paper, we provide a framework and a metric for evaluating the creativity in network steganography research.
In line with creativity research, creativity comprises the novelty and applicability of products (solutions).
Based on our creativity framework and metric, the contributions of this paper are:

\begin{itemize}%
  \item \textit{Long-term improvement of terminology}: Our framework deals with the problem of terminological inconsistencies by providing a step-by-step approach on the basis of the hiding patterns presented in \citep{Wendzel:CSUR}. A key aspect in this regard is the handling of redundant hiding techniques.
  
  \item \textit{Creativity evaluation for hiding techniques}: Our framework contributes to two aspects of the academic peer review process. Firstly, it helps authors to clearly underpin the creativity, i.e. novelty and applicability, of their proposed steganographic hiding techniques. Secondly, it helps reviewers to evaluate the creativity of steganographic hiding techniques. Therefore, the framework introduces a novelty metric. %
  
  \item \textit{Applicability in practice}: The proposed framework is designed to fit the needs of the actual workflows in academia and is thus embedded into the academic peer review process rather than being a theoretical discussion.
  
  \item \textit{Support for novel hiding techniques}: Moreover, our framework fosters the creation of entirely novel, highly creative hiding techniques by giving the most creative researchers the chance to publish new patterns.
\end{itemize}

The remainder of this paper is structured as follows. Section~2 discusses related work and draws the link between creativity research and network steganography; in addition, it provides a background on patterns and presents lessons of the pattern community to be taken into account for our work. Section~3 presents our pattern-based creativity framework while section~4 introduces the metric to evaluate the creativity of a hiding technique. Section~5 provides an exemplary walk-through for the creativity framework. We provide a discussion of our framework in section~6 and conclude in Section~7.

\section{Background}

We first discuss related work and afterwards describe the link we draw between creativity and network steganography followed by the link between patterns and network steganography. Thereafter, we describe the lessons learned from the pattern community.

\subsection{Related Work}

Terminology of computer security was discussed by a number of authors, \citep[e.g.][]{brinkley1995concepts,InfSpektrTerminology}. The terminology of information hiding and its sub-discipline, network steganography, was discussed in additional publications \citep[see][] {petitcolas1999information,CCSurveyZander,NetStegSurvey}.
Although a systematic categorization of hiding techniques is presented, these publications on network steganography lack the discussion of underlying patterns.
\citep{Wendzel:CSUR} and \citep{NetStegBook} provide the only surveying content on network information hiding using patterns --- these publications serve as basis for our work but it is not required to be familiar with these publications to understand this paper.

Our aim is not to provide a new approach for knowledge management, as a plethora of work on especially organizational knowledge management and frameworks is already available \citep{KnowledgeMgmtArticle}.
Our work is also not the first that applies patterns for knowledge management. Henninger and Corr\^{e}a discussed advantages and drawbacks of a pattern-based knowledge management already in \citep{Henninger:PatternPracticesTR} and we highlight their relevant outcomes in \prettyref{sec:LearningFromPatCommty}.
Other approaches for scientific knowledge and terminology management are especially
	\emph{i)} publication of ideas in textual form, without using the structure of patterns; these publications include books, journal articles, conference papers, technical reports, Wikis and other forms, and
	\emph{ii)} publishing ideas in structured form, e.g. in databases.
These approaches are also useful for a framework for creativity evaluation but due to the availability of hiding patterns of \citep{Wendzel:CSUR} and \citep{NetStegBook} and the high level of knowledge on patterns within the computer science community, we have chosen patterns as a basis for our framework.

However, our work is the first that is tailored to match the requirements of network steganography research by providing a metric to evaluate and compare the novelty and applicability of hiding techniques while also enabling a unified terminology and providing an integration into the peer review process.

\subsection{Bridging Creativity and Network Steganography}

Described briefly, creativity is `adaptive originality', i.e. creativity requires the generation of an idea that is both original as well as adaptive to a particular context \citep{Simonton:PosPsych:CreativityCh,Amabile83,Csikszentmihaly96,DrazinGlynnKaza99,FarmerTierneyKung03}.
Creativity is seen as one of society's biggest assets \citep{Simonton:PosPsych:CreativityCh}. Overlapping the fields of science, technology, economics and arts, creative efforts lead to competitive advantages, innovative products and processes and are honored by awards like the Nobel prize \citep{AgarsKaufmannDeaneSmith12,CaroffLubart12}.

Being such an important driver for social and economic progress and wealth, it is obvious that a great deal of research has focused on the `creativity phenomenon' \citep{MumfordGustafson88,Runco06,Ward04}. Different sciences are dealing with the topic, such as engineering, sociology, and psychology to only name the most relevant for our purpose. Due to varying research backgrounds and different research focuses and aims, conceptualizations of creativity (e.g.\ creativity as logic, genius, chance, and zeitgeist) differ and so do the findings. Smith counts more than 100 definitions of creativity \citep{Smith05}, and his focus is limited to psychological literature only. Thereby it is not surprising, that Simonton emphasizes the inconsistencies of research on creativity \citep{Simonton:CreativityScienceBook}.

However, the diverse approaches to creativity can be categorized by their primary focus. In this tradition the `\emph{4p}' classification is used: creativity as \emph{product, person, process} or \emph{press} \citep{Mooney63,Rhodes61}:

\begin{itemize}
   \item \textit{Product}: As mentioned above, for a product to be labeled creative, two characteristics are indispensable: it has to be original as well as adaptive. Original refers to: new, unusual, unique, new viewpoints, varied, breaking from existing patterns. Adaptive means: useful, valuable, effective, efficient, and contributing to society \citep{PalmerCesingerGelleriEtAl:inpress}.
  \item \textit{Person}: In psychology a strong focus lies on the identification of individual traits fostering creativity. A combination of cognitive (intelligence, knowledge), non-cognitive (personality) and motivational (need for creativity, interests, achievement motivation) abilities and predispositions determines creative potential on an individual basis \citep{Palmer:inpress}.
  \item \textit{Process}: Especially in highly experience- or knowledge-driven domains, creative ideas and innovative solutions do not emerge out of a sudden. Process research deals with the challenges one faces on his %
way from initial recognition of a problem or demand to a sustainable and accepted creative solution and describes the relevant traits and behavior in respective process phases. Psychological literature proposes several process models (for an overview see \citep{Palmer:inpress,HowardCulleyDekon08}). An initial understanding of the creative process distinguishes four process stages: preparation, illumination, verification, and implementation \citep{Lubart01}. 
  \item \textit{Press}: `Press' comprises the creative environment. This research stream investigates the variety of situational and social contextual factors influencing the creative success, such as motivation, scope of action, leadership style and required support. For an overview of influencing variables in an organizational setting, see \citep{Krause13}, \citep{OldhamCummings96}, \citep{SchulerGoerlich07} or \citep{ZhouShalley03}.
\end{itemize}

Network steganographic research describes, develops, and evaluates hiding techniques and thereby focuses on \emph{products} according to the \emph{4p taxonomy}. In the field of network steganography, many research is available, i.e. a high total output due to the number of publications is present. And, of course, the amount of publications as well as their quality (e.g. measured via citation index) can be used to quantify the impact of a person's productions \citep{Simonton:PosPsych:CreativityCh} --- measures also applied to \emph{rank} researchers. In this paper, we primarily focus on the creative products rather than on the creative person, the creative process or the creative press.

As mentioned, a large extent of publications on hiding techniques provides a rather small novelty as they belong to the same hiding patterns. To introduce a bridge between creativity and network steganography, novelty and applicability as key characteristics of creative products must be taken into account. We present a metric to evaluate both aspects.

\subsection{Bridging Patterns and Network Steganography}

Patterns provide a solution to a problem in a given context and are well-known from other scientific areas, such as software engineering \citep{PatternBuch2,KnowledgeMgmtWPatterns,PatternBuch1}. They originate from a non-computer science field, namely architecture, but are now also common in the area of computer security \citep{yoshioka2008survey}. In computer security, patterns are especially used for security engineering; they are separated into those used for requirements engineering, for the design phase and for the implementation phase of a security system.

So far, patterns were only applied twice in network steganography as a means to create a taxonomy and survey of network hiding techniques \citep{Wendzel:CSUR,NetStegBook}. In network steganography, the general problem is to hide information within the context of a network transmission; however, many solutions for this problem-context pair exist, which results in a number of patterns.

Patterns are linked to a several advantages, including their flexibility, the fact that they can serve as an easy basis for the expression and discussion of ideas, and their easy structure \citep{KnowledgeMgmtWPatterns}. These advantages make patterns a known tool for knowledge management, also outside of science in business and governmental organizations \citep{KnowledgeMgmtWPatterns}.

In general, patterns can be described in different forms, which complicate their comparison and understanding. For this reason, common \emph{languages} were developed to describe patterns in a unified manner. A well-known and established pattern language is the \emph{pattern language markup language} (PLML) \citep{plml}. All pattern languages comprise own attributes used to describe a pattern. In PLML, a pattern comprises a number of XML-based attributes, of which only a few are of significance for hiding methods \citep{Wendzel:CSUR}:

\begin{itemize}
 \item \emph{Pattern Id}: identifies a pattern
 \item \emph{Name}: the name of a pattern
 \item \emph{Alias}: alternative names for a pattern
 \item \emph{Context}: %
 		description of where a pattern is located within the hierarchy of patterns (e.g. as a  sub-pattern of another pattern)
 \item \emph{Solution} and \emph{Implementation}: describe the functioning of a pattern and add implementational details, e.g. code fragments
 \item \emph{Evidence} and \emph{Literature}: describe exemplary areas of application for the pattern and reference publications
\end{itemize}

Using such pattern languages, patterns can be easily grouped in a \emph{pattern collection}. A pattern collection comprises multiple patterns belonging to a similar domain, e.g. all patterns that describe hiding methods for networks.

As stated by Henninger \emph{et al.}, one of the main intentions of patterns is to \emph{provide a common vocabulary by which people can succinctly communicate well-known solutions to recurring problems} \citep{Henninger:PatternPracticesTR}. A so-called \emph{pattern collection} is a \emph{set of patterns addressing a fairly cohesive problem domain} \citep{Henninger:PatternPracticesTR}. Patterns for the problem domain of hiding techniques %
can thus be considered a pattern collection.

\subsection{Learning from the Software Patterns Community}\label{sec:LearningFromPatCommty}

While pattern collections offer a clear advantage of cleaning up a domain and helping to share \emph{ideas and abstractions between people working in the same conceptual space} \citep{KnowledgeMgmtWPatterns}, they are also linked to a number of challenges \citep{Henninger:PatternPracticesTR}.

Firstly, pattern collections are stored in a multitude of locations, e.g. conference proceedings or on the web (in wikis and websites of different format). For this reason, patterns are not easily accessible as many potential sources must be used to find new patterns.
Secondly, pattern collections are not easy to link as different styles for pattern collections exist, which describe patterns with a different set of attributes.
Thirdly, duplicates of patterns can arise under different names. On the other hand, Henninger mentions the need for duplicates as \emph{people may want to express the patterns different[ly]}, and patterns should allow \emph{a certain degree of expression} \citep{Henninger:PatternPracticesTR}.
Our creativity framework addresses all three issues as we will explain in the next section.

\section{A Pattern-based Framework for Network Steganography}

We first illustrate the requirements of our creativity framework and then present the framework itself by explaining all the steps that must be performed within it.

\subsection{Requirements}

Our aim is to learn from the known drawbacks of pattern collections which we discussed in the previous section and to adjust our framework to scientific practice. Hence, we compiled the requirements to address these drawbacks in our creativity framework:

\begin{enumerate}
\item The framework must require a researcher to publish his patterns in a publicly accessible way to ensure that all patterns and all updates to the pattern collection are accessible to the scientific community.
\item The framework must emphasize the need for a unified form of pattern description, which is the \emph{pattern language markup language} (PLML).
\item The framework must address the aspect of duplicates; PLML contains an attribute that allows aliases for existing patterns.
\item To find use in practice, the framework must comply with the common workflow of scientific research, especially within a peer review.
\item The framework must not rely on the participation of all scientists of the research community. It is unlikely that a whole community will accept and work according to the same framework as people may not even aware of the framework or reject its idea. 
\item The framework should foster the development of new and applicable, that is creative, hiding methods.
\end{enumerate}

\subsection{Creativity Framework}

\begin{figure*}[!th]
\centering
\epsfig{file=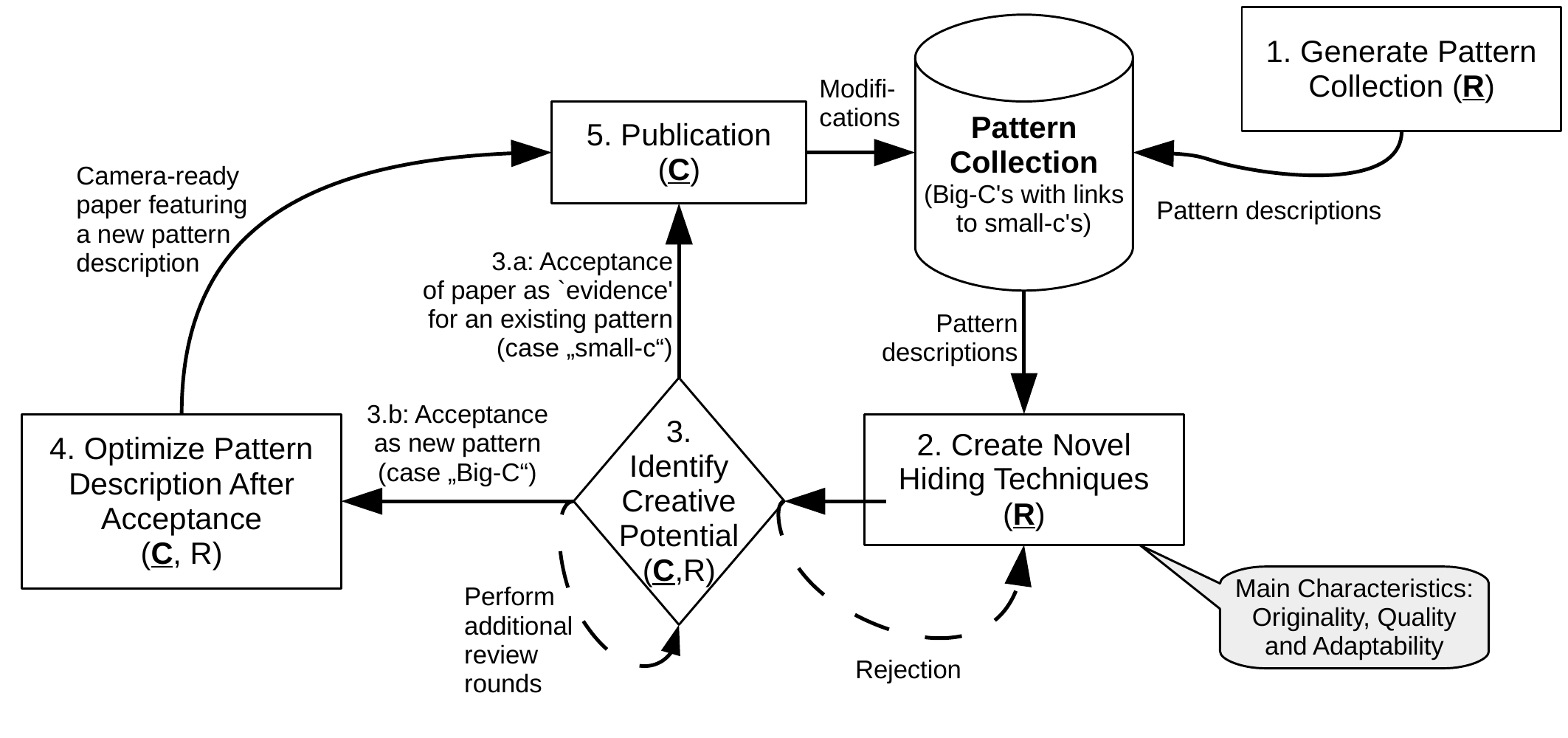, width=1\textwidth}
\caption{Creativity framework for network steganography}
\label{fig:Framework}
\end{figure*}

Our creativity framework is visualized in \prettyref{fig:Framework} and consists of five steps, which we will explain one after another. Each step in the framework is performed by at least one role; either the scientific community (C), which performs peer reviews and maintains the pattern collection, or the researcher(s) (R), which publish novel network steganographic techniques. If a step of the framework is performed by both roles (C and R), the leading role is highlighted in \prettyref{fig:Framework}.

\subsubsection{Step One: Generate Pattern Collection}

Before a pattern collection can be used and serve as basis for further research, it must be created by one or many researchers (R). This step is performed by compiling a survey in which recurring design principles of hiding techniques are analyzed and grouped into hiding patterns. These patterns must be explained in the form of a pattern language -- in our case PLML -- to provide a unified and clear structure of the grouped hiding techniques. The compilation process for a new pattern collection may require several months of work. By publication in a public journal, the pattern collection becomes available to the scientific community (C).

The hiding patterns described in \citep{Wendzel:CSUR,NetStegBook} are suitable for the initial pattern collection. Due to the detailed description of the approach, these publications %
can also serve as a guideline for the development of new pattern collections.

\subsubsection{Step Two: Create Novel Approaches}

Completely new and thereby highly creative ideas are quit rare. In fact, a closer look at creative contributions shows that a lot of innovative products on the product market, theories in science or contributions in literature and arts are reinterpretations or advancements of already existing concepts. For example, Kepler proposed his three laws of planetary motion after he drew an analogy between planetary observations and magnetism \citep{GentnerBremEtAl97}. %
Accordingly, creativity research focusing on the process perspective also emphasizes the importance of conclusions by analogy. Many models of the creative process include a distinct process stage often labeled \emph{category combination}. Category combination comprises the combination or reorganization of knowledge (respectively, information) for the generation of new ideas \citep{BaughmanMumford95,MumfordBaughmanEtAl97,MumfordConnellyGaddis03,MumfordGustafson88}.

Various offensive as well as defensive cyber security techniques, including firewalls, intrusion detection systems, honeypots, botnets, and worms, follow ideas whose equivalents can be found in nature \citep{SecurityLessonsFromNature15}. Not only does their technical logic represent an adaption of proven evolutionary concepts, but also the naming of these techniques stresses their close link to already existing mechanisms.

Applying concept combination from creativity research to the area of network steganography, researchers (R) should thus keep the patterns of the pattern collection in mind when designing new hiding techniques. For instance, an existing pattern's technique could be inverted or combined with other patterns to build \emph{ideas} for new patterns.
Therefore, a researcher may select the most diverse patterns to produce higher novelty in comparison to extending ideas of hiding patterns with lower diversity. However, a sole parallel application of two patterns (e.g. a timing channel pattern combined with a storage channel pattern) cannot be considered a \emph{new} pattern as each technique has to be handled by a separate pattern. It may, however, be imaginable to introduce hybrid patterns which combine existing patterns in useful ways. Such hybrid patterns must be considered as second tier patterns and cannot possess the same status like non-hybrid pattern as firstly, basically all patterns can be combined to form hybrid patterns, and secondly, the idea of hybrid hiding methods was already published \citep[by][]{hybrid:LACK}.

Indeed, a researcher may apply his very own discovery methods and can create entirely novel approaches without taking existing patterns as a basis. An uncommon approach to discover novel patterns may even result in higher creativity \citep{Simonton:CreativityScienceBook}. In this case of applying own discovery methods, the researcher may only use the pattern collection to ensure that his new hiding technique is not already represented by an existing pattern. In other words, the framework does not limit the creativity and freedom of researchers but aims to support it.

However, all proposed hiding techniques of a researcher must fulfill the requirements of a creative product. Firstly, the new techniques must be original with respect to the existing patterns; secondly, they must be adaptive, i.e. applicable in practice, and must be proven by an implementation.

\subsubsection{Step Three: Identify Creative Potential}

In the form of a scientific paper that features a PLML-based description of the hiding technique, the researchers (R) submit their pattern (or evidence for an existing pattern) for peer review to the research community (C).

Both, C and R, evaluate the creativity of the proposed technique, while R initiates the process. Firstly, R evaluates his hiding technique (the creative product) based on his \emph{own internal criteria for what can be considered a promising idea} \citep{Simonton:CreativityScienceBook}. If R concludes that his idea matches the necessary quality requirements for a scientific contribution, he needs to prepare a scientific manuscript for submission. In the manuscript, he underpins the creativity of his new technique using the creativity metric we introduce in \prettyref{sec:NoveltyMetric}. Afterwards, R submits the manuscript for peer review.

The peers (C) rate whether the stated arguments of R are actually true by performing a review, e.g. as described in \citep{TaskOfAReview}. Within the review, the peers additionally evaluate the creativity of R's work by applying the same metric of \prettyref{sec:NoveltyMetric}.

This metric will help to distinguish between contributions to network steganography research of a high level of creativity and such contributions that improve our understanding of steganographic techniques but do not widen the existing pattern collection. The classification of new ideas, products or concepts by their level of creativity is quite popular in creativity research.
In Psychology, the term `small-c' is used to express that a creative product is linked to low creativity \citep{Simonton14More,SilviaEtAl14,KaufmanBeghetto09,KaufmanBeghetto13}. 
In such a case, the work represents an \emph{existing} pattern and can be added to the references of the particular pattern, i.e. R's proposed hiding technique is added to the pattern collection in order to provide \emph{evidence} for an existing pattern. Therefore, R must (monitored by C's peer review) explicitly reference the existing pattern, which leads to step five. Indeed, R is able to withdraw and re-submit his work to another conference or journal if he does not accept the small-c categorization.\footnote{While most journals allow to have multiple peer review rounds per submission, conferences often directly accept or reject a submission. For this reason, conference reviewers (C) must explicitly state that a paper should be accepted under the condition that \emph{i)} R references a particular pattern and that \emph{ii)} R declares his contribution as `evidence', what should be monitored by the conference chairs (also C) as the reviewers will have no further control over the process.} %

In case of a high level of creativity, the term `Big-C' is used. In such a case, a new pattern can be created and added to the collection, which leads to step four. In both cases, `small-c' and `Big-C', the publication of a paper containing a pattern (or adding `evidence' to an existing pattern) automatically integrates R's contribution to the pattern collection.

\newpage
Note that C can also reject the work of R, which means that no alternation of the pattern collection is achieved. Criteria for rating the creativity level of R's hiding technique are presented in \prettyref{sec:NoveltyMetric}.

\subsubsection{Step Four: Optimize Pattern Description After Acceptance}

In step four, which only applies for `Big-C' cases, R's idea was accepted as new pattern and will be published. To this end, R optimizes the pattern description within the submitted paper based on C's review. This step is a part of the process in which R creates the camera-ready version of his paper.

\subsubsection{Step Five: Publication (Maintenance of the Pattern Collection)}

Within the network steganography community, research groups -- forming the people behind the role `C' -- are maintaining the pattern collection, basically due to peer review in step three. A project website for long-term maintenance of network steganography patterns was set up that allows participants to discuss and question existing patterns and their evidence: \textit{ih-patterns.blogspot.com}. Every author showing an accepted paper that adds a pattern or evidence to an existing pattern can request that his publication will be added to the webpage. This allows an easy access for the research community to the pattern collection.

However, a scientific research field may not only be driven by the power of few established research groups. It is thus necessary that a pattern collection is not only accessible due to scientific publications, but also \emph{forkable}. Forking a pattern collection is similar to forking an open source software project. In the case of an open source project fork, the code is copied; the existing contributor's names remain but the copy of the code is from that date on edited independently from the original and the names of the new contributors are added to the code. In other words, the pattern collection can be copied and extended by other individuals which can be part of C or stay even outside of the previous C. This allows changing the rules applied to evaluate the creativity of the hiding methods and also the rules of the framework. However, the reputation of a pattern collection is represented by its acceptance by leading scientists in a field and by the publication in high-quality journals and conference proceedings.

One aspect that can lead to a fork are so-called \emph{multiples}: Simonton highlights multiples which appear on a regular basis in scientific history in form of parallel discoveries or rediscoveries \citep{Simonton:CreativityScienceBook}. As mentioned, such multiples -- or \emph{duplicates} -- also appear in the area of patterns \citep{Henninger:PatternPracticesTR} but Simonton adds the aspect of \emph{grade} to it \citep{Simonton:CreativityScienceBook}. The grade is the \emph{number of rival claimants to the discovery or invention}. For instance, if $n$ authors propose the same hiding technique, the associated pattern multiple's grade is also $n$. The scientific community, which is in charge of the pattern maintenance, takes care of spotting multiples and their divergent naming.\footnote{It is important to emphasize the fact that creativity research differs between the origination and acceptance of ideas \citep{Simonton:CreativityScienceBook,HammondNeffFarr11,Lubart01}
The research community may forget non-accepted research work over the years, which can lead to a \emph{rediscovery} of the same idea by another researcher.} Therefore, pattern duplicates can be added as \emph{aliases} to an existing pattern (step three) and members of C can act as R to propose such changes (step two).

\section{A Network Steganography Creativity Metric}\label{sec:NoveltyMetric}

One of the core goals of our framework is to provide a metric for the creativity of a hiding technique. Such a metric cannot be built on a single aspect as hiding techniques in network steganography are linked to a variety of attributes.

Therefore, we provide a threefold metric, which requires a textual description within R's manuscript for each of the following categories. Of these three categories, the primary category is considered the most important and the tertiary category the least important to evaluate the creativity of a hiding method:

\begin{enumerate}
 \item Primary category (\textit{originality of hiding technique}): The major aspect to evaluate the creativity of a hiding technique is the extent to which it differs from the existing hiding techniques represented in known hiding patterns. Due to the large divergence of hiding techniques, a textual representation is necessary to explain this part. For instance, using a reserved flag of a network protocol cannot be considered novel as patterns already describe  this technique.
 
 \item Secondary category (\textit{steganographic quality}): A researcher (R) can take various steganographic attributes into account to support the quality of his hiding technique. The classical aspects to highlight in this regard are the detectability, robustness, and bandwidth of a hiding technique, but it is also possible to highlight its steganographic cost \citep{StegoCostArticle}. An optimally prepared manuscript highlights all four attributes.
 
 \item Tertiary category (\textit{adaptability} or \textit{novelty of application area}): A steganographic hiding method may be applied to a new area (e.g.\ to smart vehicle networks) that was not subject to a network steganographic research before. A new area of application can also be a particular network protocol. However, a new area of application does not necessarily increase the creativity of the hiding method itself since it represents only the adaptation of an exiting idea to another context.
Instead, the adaptation to a new area supports the addition of R's hiding technique to the `evidence' attribute of a given hiding pattern.

In other words, the novelty of the application area depends on the time of a technique's presentation. For instance, several years ago, steganographic methods in smart buildings might have been a novel application area but after first publications arose in this context, the novelty of the application area is now lower. Similarly, hiding information in the IP header was a newer area of application in the mid-1990's than it is today.
\end{enumerate}

In step three of the framework, the researcher (R) provides arguments in his paper to support all three categories while the reviewers (C) review the correctness and reasonability of the provided arguments. On this basis, C can decide whether the approach is a new pattern, `evidence' for an existing pattern, or not novel enough or of not acceptable quality to become a part of a pattern collection. If C decides to allow R a modified re-submission of his work, an additional review round can be performed --- possibly multiple times.

\section{Exemplary Walk-though}

For a better illustration, we now provide an example on how to use the creativity framework. We assume that both, R and C are aware of our framework. If neither R nor C is aware of the framework, the framework would simply remain unused by these individuals, providing no update to the pattern collection at all. However, publications not based on patterns can be integrated into pattern collections a posteriori by creating surveys or by adding evidence to online pattern collections.

Step one of the framework (the creation of a pattern collection) must only be performed if no pattern collection is already available.
For this reason, R can use the existing pattern collection as the basis for the following steps.

In step two of the framework, R combines the concepts of a reasonable number of patterns in the hope to create a new pattern --- as mentioned, a typical process to generate a creative product. Alternatively, R can create ideas for new hiding techniques from scratch without taking existing patterns into account at this step.

Finally, R ends up with a considerably useful idea, namely to signal hidden information by the `position' of a packet corruption within a network flow. For instance, if the third packet in a network flow is intentionally corrupted, it represents a hidden symbol $A$ while a corruption of the sixth packet represents a hidden symbol $B$.
 
Being aware of the framework, R can recognize that his hiding technique is a combination of the two patterns \emph{PDU Corruption/Loss} and \emph{Position}.
The PDU Corruption/Loss pattern represents hiding techniques which signal hidden data via packet loss or via corrupted network packets (e.g.\ those having invalid checksums).
The Position pattern transfers hidden information by modifying the order of header elements (e.g.\ the order of header lines of in a HTTP request).

Based on his idea of a hiding technique, R implements a prototype to verify the feasibility of the technique and to evaluate its technical details such as the channel capacity. R compiles all relevant information about his hiding technique in form of a scientific paper. When describing the hiding technique in his paper, R either explains to which patterns the technique provides new evidence or why the technique cannot be associated with an existing pattern and, for this reason, necessarily represents a new pattern. Finally, R submits his paper for peer review to a scientific conference.

During the review process, R's paper is sent to several peer reviewers (C). The peer reviewers state that the major aspect of the proposed hiding pattern is the position of a given PDU in a network flow (in R's case, the important PDU is the one that contains an error). Therefore, the reviewers consider the hiding technique as being `small-c', i.e. it provides additional `evidence' to the Position pattern, but cannot be seen as a new pattern itself (still step three).

Based on all received comments of the reviewers, R finalizes his paper and submits a camera-ready version for publication. After the camera-ready version of the paper is published and if patterns were taken into account, the `evidence' is officially provided to the existing pattern through the accessible paper (step five).

If R's proposed hiding technique would have been accepted as a new hiding pattern (Big-C) by the peer reviewers, the published paper would represent the description of a new pattern, eventually featuring an optimized description to match the reviewer's requirements (step four).

R can support the visibility of his hiding technique (being it a patern or an evidence entry to an existing pattern) in form of a comment on the existing pattern listings (e.g. \emph{ih-patterns.blogspot.com}) or at any other place where a pattern collection is published to increase the distribution of research findings.

\section{Discussion}

To achieve our goal of a practical framework, we do not provide a low-level discussion on scientific creativity as can be found in \citep{Simonton:CreativityScienceBook}. For instance, our framework does not focus on the slight, and in some aspects unclear, differences between multiples and their opposite (singletons). If the scientific community decides to accept the proposed framework and makes the decision to change the aspects of it or to add the handling of more details of scientific creativity, the model is open for change.

We will first discuss whether our framework achieves the previously introduced framework requirements, followed by a description on how to match the general requirement that a pattern cannot be called a \emph{pattern} until at least three evidence cases are provided. We end this section with a note on alternative application areas for the framework.

\subsection{Discussion of the framework's requirements}

We address the requirement of publicly accessible patterns (requirement one) by requiring R to publish novel patterns in publicly accessible, peer reviewed organs and by providing the option to provide comments to the existing pattern collection on-line or on alternative websites every researcher can create himself.

The requirement of a unified pattern description (requirement two) is addressed by the use of PLML as it ensures unified descriptions by its pre-defined set of attributes (e.g. evidence, alias or solution).

Our framework limits duplicates (requirement three) due to the framework's integration into the academic peer review process in which duplicates may be spotted. In addition, requirement three is addressed by the use of PLML aliases. Aliases allow the a posteriori merging of redundant ideas published using different names, i.e. they can be applied if duplicates were not identified during the peer review.

The integration into the peer review process matches the requirement of the framework's applicability to scientific practice (requirement four). Given the framework awareness of R or C, the consideration of patterns for a new hiding technique can be enforced. At the same time, the traditional review procedures are kept. The overhead of checking whether a proposed hiding technique matches an exiting pattern is minimal due to the small number of patterns and their hierarchical order.

The framework is applicable even under the circumstance that only a minority of researchers of the network steganography domain will apply it (requirement five). Even if only applied by few researchers, their combined effort for a unified pattern collection will lead to a more unified terminology and less re-inventions for hiding techniques. However, the more researchers apply in the framework, the more efficient it will be. In that sense, the framework represents a living process and the number of researchers participating in it can change over time.

Our framework fosters the creation of novel hiding techniques (requirement six) by giving researchers the chance to contribute an own pattern to the pattern collection. On the other hand, our concept can lead to the situation that a proposed hiding technique becomes a pattern although it should only be considered as `evidence' for an existing pattern. For instance, one researcher could propose a new pattern and an unqualified reviewer might accept it as such although it should only be accepted as an `evidence' entry for an existing pattern (or not at all) --- this contradicts requirement three and is an existing problem of the network steganographic field that the framework will not solve but reduce if applied correctly.

Creativity research focusing on creative products outlines the problems related to creativity ratings of products like hiding techniques. Whenever products are rated, this evaluation underlies a social construction of criteria. Which solution is considered as non-, lowly or highly creative depends strongly on the raters' background. Their individual expertise, experience and strategy influence ratings as well as the zeitgeist, cultural factors or simply a changing technological context in change of time. Little expertise or experience of the raters is one of the key issues. Rather uninformed reviewers might not be aware of already existing techniques. In consequence, they might not be able to detect the linkage of an alleged new pattern to an already known pattern of hiding techniques.

To deal with such problems and the resulting inconsistencies, we recommend compiling pattern surveys every 3-10 years and publishing these surveys in recognized journals or conferences. New surveys can modify the pattern collection by publishing a new version of it. Such surveys can, for instance, move a `pattern' into the `evidence' attribute of an existing pattern. The proposal of regular survey compilation matches the requirements four and five for integration into scientific practice as it keeps the pattern collection updated and provides good knowledge management for the scientific community as well as for practitioners. In addition, a scientific discussion about patterns should be performed using moderated websites as these can be updated at any time while surveys remain as mid-term and long-term solutions.

\subsection{Pattern's Requirement of Recurring Designs}

Patterns must -- per definition -- occur multiple times. A typical boundary value for a design to become a pattern is three occurrences, provided as references in the `evidence' attribute of PLML. It is our belief that the scientific community should ensure that all hiding techniques with `Big-C' are represented by new patterns and rediscoveries are kept at a minimum. For this reason, we suggest that for the few new patterns to be found, patterns with a `pending' status are created. Such patterns are part of a pattern collection as all other patterns but feature the keyword `pending' in their name. A simple heuristic could be to consider the `pending' keyword obsolete as soon as three references are listed in the `evidence' attribute. An actual removal of the keyword cannot be done a posteriori in the same publication but by the above-mentioned regular surveys.

\subsection{Applicability in Other Areas}

We assume that our creativity framework can also be applied to other areas of information hiding besides network steganography. An imaginable area is digital media steganography. Moreover, our framework can be used to create a pattern collection of steganographic countermeasures, which were also already linked to patterns in \citep{Wendzel:CSUR}. For instance, various techniques similar to the \emph{pump} \citep{KangMoskowitz:Pump,CC_Elimination_Protos} can form a pattern while \emph{traffic normalization} \citep{NormalisierungsPaper} must be considered a clearly different pattern due to its fundamentally different functioning.
The initial step of creating a pattern collection must be performed first in these areas.
Finally, the framework will be applicable to non-information hiding, even non-computer science, areas due to its generic approach.

\section{Conclusion}

We provide a framework and a metric for evaluating the creativity linked to hiding techniques and to handle inconsistencies in the terminology of network steganography. The framework can be applied in practice and is thus embedded into the academic peer review process. By providing an exemplary walk-through, we have shown the applicability of the approach. The framework's design is not static and can be modified by the scientific community to fit their needs and it can be adapted to future developments. As patterns serve as basic elements of our framework, the framework takes advantage of their flexibility, accessibility and structure. In addition, the community benefits from the framework's application even if only applied by a minority of researchers.

A drawback of our framework lies in the fact that different, especially low-qualified, researchers may allow the integration of a pattern that is only a representation of another hiding technique. We foresee regular surveys (every 3-10 years) as a clean-up solution for this drawback.

In addition to steganographic hiding techniques, the creativity framework is also applicable to other areas of information hiding, such as countermeasures or digital media steganography, as well as to other sciences.

\section*{Acknowledgements}

We highly appreciate the valulable comments received by the reviewers and we like to thank Jernej Tonejc and Jaspreet Kaur for proof-reading our paper.

\bibliographystyle{jucs} 
\bibliography{arxiv} 

%

\end{document}